\documentclass{llncs}
\usepackage[utf8]{inputenc}
\usepackage{url}
\usepackage{framed}
\usepackage{amsmath,amsfonts,amssymb,amstext,enumerate}
\usepackage{color}
\usepackage{xspace}
\usepackage{varioref}
\usepackage{hyperref}
\hypersetup{
  linkcolor  = red,
  citecolor  = green,
  urlcolor   = blue,
  colorlinks = true,
  breaklinks=true,
  bookmarksopen=true
}
\usepackage[nameinlink,capitalize]{cleveref}
\usepackage[
	operators,
	sets,
	adversary,
	landau,
	probability,
	notions,
	logic,
	ff,
	mm,
	advantage,
	primitives,
	events,
	complexity,
	asymptotics,
	keys,
	lambda]{cryptocode}

\title{Private Electronic Payments with Self-Custody and Zero-Knowledge Verified Reissuance}

\author{Daniele Friolo \and Geoffrey Goodell
\and Dann Toliver \and Hazem Danny Nakib}
\institute{}
\date{}

\newcommand{\Sim}{\mathsf{S}}

\newcommand{\advD}{\mathsf{D}}

\newcommand{\getsr}{{\:{\leftarrow{\hspace*{-3pt}\raisebox{.75pt}{$\scriptscriptstyle\$$}}}\:}}

\newcommand{\rel}{\Rel}

\newcommand{\wit}{\ensuremath{w}}

\newcommand{\Gen}{\ensuremath{\mathsf{Gen}}}
\newcommand{\Commit}{\ensuremath{\mathsf{Com}}}

\newcommand{\hyb}{\ensuremath{\mathbf{Hyb}}}

\newcommand{\chal}{\ensuremath{\mathcal{C}}}

\newtheorem{Claim}{Claim}

\newcommand{\calQ}{\mathcal{Q}}

\newcommand{\calU}{\mathcal{U}}

\newcommand{\msg}{m}
\newcommand{\com}{\mathit{\gamma}}

\def\bitcoinB{\leavevmode
	{\setbox0=\hbox{\textsf{B}}%
		\dimen0\ht0 \advance\dimen0 0.2ex
		\ooalign{\hfil \box0\hfil\cr
			\hfil\vrule height \dimen0 depth.2ex\hfil\cr
		}%
	}%
}

\newcommand{\commit}{\ensuremath{\mathsf{Commit}}}

\newcommand{\ver}{\ensuremath{\mathsf{Vrfy}}}

\newcommand{\ignore}[1]{}

\newcommand{\users}{\calU}

\newcommand{\minter}{T}

\newcommand{\banks}{\mathcal{B}}
\newcommand{\cbank}{CB}

\newcommand{\bank}{B}
\newcommand{\user}{U}

\newcommand{\token}{F}

\newcommand{\receiver}{R}
\newcommand{\sender}{S}

\newcommand{\BS}{\mathsf{BS}}
\newcommand{\blind}{\mathsf{Blind}}
\newcommand{\unblind}{\mathsf{Unblind}}
\newcommand{\blinded}{\beta}
\newcommand{\bfactor}{\tau}
\newcommand{\checkreg}{\mathsf{CheckRegulation}}

\newcommand{\post}{\mathsf{``post"}}
\newcommand{\readcom}{\mathsf{``read"}}

\newcommand{\stm}{x}
\newcommand{\prove}{\mathsf{Prove}}
\newcommand{\setup}{\mathsf{Setup}}
\newcommand{\crs}{\mathsf{crs}}
\newcommand{\game}{\mathbf{Exp}}
\newcommand{\Adv}{\mathsf{A}}
\renewcommand{\adv}{\mathsf{A}}

\renewcommand{\sig}{\mathsf{Sign}}

\newcommand{\bsign}{\mathsf{BlindSign}}
\newcommand{\bsigma}{\tilde{\sigma}}

\newcommand{\msgspace}{\mathcal{M}}
\newcommand{\comspace}{\mathcal{C}}
\newcommand{\sigspace}{\mathcal{S}}

\newcommand{\STORED}{\mathsf{``stored"}}
\newcommand{\denied}{\mathsf{``denied"}}

\newcommand{\Rel}{\mathcal{R}}

\newcommand{\pidc}{\pi_\mathsf{DC}}

\newcommand{\bEUFCMA}{\mathsf{bEUF}\textrm{-}\mathsf{CMA}}

\newcommand{\sfQuery}{\mathsf{q}}
\newcommand{\tsigma}{\tilde{\sigma}}

\newcommand{\vkcb}{\vk_{\cbank}}
\newcommand{\skcb}{\sk_{\cbank}}

\newcommand{\oracle}{\mathcal{O}}

\newcommand{\SIG}{\mathsf{SIG}}

\newcommand{\corr}{\mathcal{I}}

\newcommand{\msgs}{M}

\newcommand{\bboard}{BB}

\newcommand{\stored}{\STORED}

\newcommand{\vkrz}{\vk^{\receiver_0}}

\newcommand{\vkro}{\vk^{\receiver_1}}

\newcommand{\vksz}{\vk^{\sender_0}}

\newcommand{\vkso}{\vk^{\sender_1}}

\newcommand{\sigsz}{\sigma^{\sender_0}}
\newcommand{\sigso}{\sigma^{\sender_1}}

\newcommand{\skrz}{\sk^{\receiver_0}}

\newcommand{\skro}{\sk^{\receiver_1}}

\newcommand{\skso}{\sk^{\sender_1}}

\newcommand{\bankso}{\bank^{\sender_1}}

\newcommand{\blindedso}{\blinded^{\sender_1}}

\newcommand{\vkspace}{\mathcal{VK}}

\newcommand{\indexes}{\mathcal{I}}

\newcommand{\verassets}{\msgs_{valid}}

\newcommand{\Reldc}{\Rel_\mathsf{DC}}

\newcommand{\validity}{\mathsf{IsValid}}

\newcommand{\getburnt}{\mathsf{GetBurnt}}

\newcommand{\msgsburnt}{\msgs_{burnt}}

\newcommand{\restrict}[2]{#1^{#2\rceil}}

\newcommand{\td}{\mathsf{td}}
\newcommand{\Poly}{\mathsf{poly}}

\begin{document}
\setlength\parskip{1em}

\maketitle

\begin{abstract}
This article builds upon the protocol for digital transfers described by
Goodell, Toliver, and Nakib (WTSC at FC '22), which combines privacy by
design for consumers with strong compliance enforcement for recipients of
payments and self-validating assets that carry their own verifiable provenance
information.  We extend the protocol to allow for the verification that
reissued assets were created in accordance with rules prohibiting the creation
of new assets by anyone but the issuer, without exposing information about the
circumstances in which the assets were created that could be used to identify
the payer.  The modified protocol combines an audit log with zero-knowledge
proofs, so that a consumer spending an asset can demonstrate that there exists
a valid entry on the audit log that is associated with the asset, without
specifying which entry it is.  This property is important as a means to allow
money to be reissued within the system without the involvement of system
operators within the zone of control of the original issuer.  Additionally, we
identify a key property of privacy-respecting electronic payments, wherein the
payer is not required to retain secrets arising from one transaction until the
following transaction, and argue that this property is essential to framing
security requirements for storage of digital assets and the risk of blackmail
or coercion as a way to exfiltrate information about payment history.  We claim
that the design of our protocol strongly protects the anonymity of payers with
respect to their payment transactions, while preventing the creation of assets
by any party other than the original issuer without destroying assets of equal
value.

\end{abstract}

\section{Introduction}
Privacy and scalability are often at odds in digital payment solutions.  We
assert that this is largely the result of the involvement of the issuer in the
circulation of electronic money.  Goodell, Toliver, and Nakib have shown that
it is possible to achieve both privacy and scalability imperatives simultaneously by combining a
privacy-respecting token issuance process with a model by which assets track
their own provenance~\cite{goodell2022}.  Unlike most real and proposed modern
retail payment solutions that are currently used in practice (public and private), including interbank payments, cryptocurrency
transfers, and centralised e-money, their approach removes the issuer from the
transaction channel, allowing transactions to take place at the edge.  However,
the protocol assumes that before value transferred in one consumer transaction
can be spent by a consumer in another transaction, the underlying assets must
be recirculated by \textit{minters}, which are entities authorised and trusted by the
original issuer to issue new assets to replace previously spent assets.  It is
assumed that minters do not produce valid assets without authorisation from the
issuer, despite having the technical capacity to do so.  Although this
assumption mirrors the mechanical control that central banks have over
private-sector companies that produce banknotes and coins, such an assumption
might be less appropriate for an environment wherein valid assets can be
created without the aid of proprietary technology and expensive materials with
large fixed costs.

We show that it is possible for a payer to demonstrate that an asset was
created in accordance with the rules governing its creation without providing
evidence linking the asset to the specific context and circumstances of its
creation.  The combination of privacy and resistance to counterfeiting is as
important to digital money as it is for physical cash.  Users of cash have a
reliable way to verify the authenticity of banknotes while also maintaining that they
would not expect that the serial numbers of their banknotes would be used to
associate the bank account of a payer who had withdrawn the banknotes from her
bank to the bank account of the recipient who had deposited the banknotes
following a retail transaction.

In this article, we describe a mechanism that combines zero-knowledge (ZK) proofs
with a robust audit log to provide a way for payers to prove that their assets
were created during an atomic process wherein assets of equal value were
destroyed, without incurring the risk of de-anonymisation or undermining the
properties claimed in the design of the original protocol.  We show that using
this new mechanism allows the process of reissuance of spent tokens to be
undertaken by relatively untrusted minters, or by consumer-facing banks instead
of minters, thus improving the robustness of the protocol relative to the
original design and reducing its attack surface.

This article is organised as follows.  In the rest of this section, we provide
background to motivate the design of our payment architecture and characterise
the requirements that underpin its design.  In the following section, we compare
related protocols and assess their strengths and weaknesses.  In
\cref{s:prot-overview}, we provide the protocol description.
  In
\cref{s:analysis}, we offer a formal security analysis of the properties our protocol aims to satisfy and the main theorem.  In \cref{s:conclusion}, we conclude,
pointing to the next steps and future directions for this line of research.
For space reasons, we background assumptions
and definitions that are foundational to our design are given in \cref{s:preliminaries}. The formal description of our protocol and the proofs for the main theorem stated in \cref{s:analysis} are deferred to
\cref{s:formal-desc} and \cref{s:proofs} respectively.

\subsection{Background}

\ignore{New and emerging developments of payment infrastructure is on the rise.
There are several considerations driving this exploration. In particular
lies the secular decline in the use of cash as a means of payment in retail transactions,
particularly in industrialised economies, over the last decade as consumers move to internet payments, including debit and credit cards. The decline of the use of cash is fundamental because the properties of cash, in particular privacy and choice, have been essential and expected by the public. Absent the privacy-preserving properties of cash, users are at constant risk of being profiled by various parties.

The value of retail payment infrastructure relative to other forms of
central bank money lies in its suitability for conducting payments.
Payments can be broken down into two categories, namely cash and electronic payment systems.

The use cases for cash often require low latency, low transaction overhead, and unlinked transactions (consumer privacy by design). A good mechanism for payments can both meet these requirements and also extend the benefits of digital transfers into the cash domain, so that the beneficial properties of both will become available in the digital realm. As the digital economy continues to expand into more facets of consumers' everyday lives, ensuring the cash-like properties such as efficiency, security and ownership is paramount.

}This article extends previous work in a category of protocols, one example of which is provided by Goodell, Toliver, and Nakib~\cite{goodell2022}, that allow assets that had been created via a Chaumian mint~\cite{chaum1982} to be transferred between independent parties without the involvement of the issuer. Important characteristics of these protocols include privacy, self-custody and compliance enforcement for recipients without relying upon exogenous punishment mechanisms which sit outside of the protocol.  Punishment mechanisms are unnecessary and introduce negative
externalities, such as reliance upon a panoptical view of the behaviours, transactions, and activities of all participants.  In addition, these protocols exhibit \textit{transaction independence} in the sense that they do not require a user to retain any secret information that might arise from a transaction.  This property is important because it means that the user can optionally expunge all records of the transaction, should they wish to.  Because payers can reasonably be expected to be unable to furnish information about payment history, the de-anonymisation risk posed by the compromise of a device, blackmail, or coercion, is mitigated relative to schemes that require payers to maintain information about previous transactions, for example, to prove, via a zero-knowledge mechanism or otherwise, that the state of a balance within the system was updated in accordance with the rules.

In particular, this new category of protocols is able to achieve the
combination of privacy by design, self-custody, and compliance enforcement by using an external mechanism for providing integrity verification of assets, wherein the assets are held outside of the relevant ledger.  Rather than managing tokens directly, the ledger instead plays the role of integrity verification for the state of the assets.  However, because privacy for consumers depends upon the inability to link an asset to the circumstances of its creation, the integrity of the assets must depend upon the inability of those charged with the task of
creating assets to do so against the rules.

This article offers an extension to protocols in this category that allow users to prove that an asset is correctly produced while not needing to establish a specific relationship to its creation.  Specifically, the mechanism presented in this article provides a way for a prover to demonstrate that an asset has been correctly created by a party whose task is to recirculate tokens (i.e., to
``burn'' and ``mint'' in equal measure) within a system, without revealing any information about its creation. Since the assets exist outside the ledger with the ledger playing the role of integrity verification for the state of the assets, it is essential for parties receiving assets to be able to establish their security properties.  By providing a way for assets to be verified without linking an asset being proved to an asset being created, it becomes possible to allow the recirculation of assets by parties that operate independently of issuers, without compromising an essential privacy characteristic for digital payment infrastructure.

The mechanism presented here is generic, in the sense that it can be applied to a variety of privacy-enhancing payment schemes that rely upon a Chaumian mint. However, the primary value of the mechanism lies in its use in cases in which assets are transferred without the involvement of their creator, so it will be useful to show how to use this mechanism with protocols that allow assets to be held directly and for which verification is based upon proofs of provenance. For this reason, in future work, we intend to prove how the mechanism works
with provenance-based transfers that use Merkle tries to ensure the integrity of assets.

\subsection{Design requirements}

Next, elaborating on the set of requirements proposed by Goodell, Toliver, and
Nakib~\cite{goodell2022}, we describe the foundational requirements for our
design to support the stated characteristics of consumer privacy and system
scalability while ensuring that the system can be adopted within the broader
socio-technical context of modern retail payments:

\begin{itemize}
    \item \textit{Consumer privacy by design.}  Consumers must be able to verify that, in their spending transactions, they do not disclose any information that can be used, \textit{ex ante} or \textit{ex post}, to link their identities to the transactions.
    \item \textit{Compatibility with regulatory objectives.}  The design of the system must be compatible with the expectation that consumers would be subject to AML/KYC checks at the point in time at which they receive tokens and that merchants would be subject to AML/KYC checks sometime between the point in time at which they receive tokens from consumers and the point in time that they deposit the tokens or reuse them in subsequent transactions.
    \item \textit{Independent transactions.}  Importantly, tokens held by consumers within the system must not be associated with accounts.  Consumers must not be expected to hold any secrets other than those related to the set of tokens that they currently hold, so there is no possibility to reconstruct or confirm the prior payment history of an honest consumer even if an adversary takes control of a user's device.
    \item \textit{No involvement of the issuer in the circulation of tokens.}  In the protocol proposed by Goodell, Toliver, and Nakib~\cite{goodell2022}, it is assumed that the issuer would not participate in transactions.  However, the issuer would still implicitly have a role in the reissuance of tokens after they are spent, because it is assumed that they must enforce the responsibility of minters to not create invalid tokens. In this version of the protocol, it must be possible for minters to be untrusted, so that unauthorised tokens can be rejected at the point of sale.
    \item \textit{Efficient procedure for reissuance of tokens.} Since tokens cannot be rehypothecated, there must be an efficient mechanism for recirculating tokens within the system that avoids requiring banks to hold too many tokens at once.
    \item \textit{Efficient proof generation.} The zero-knowledge proofs used by our system should be computable efficiently, in terms of both communication and computation, and parallelisable so that they can be efficiently performed at scale.
    \item \textit{Stateless issuer.} The Central Bank is trusted and generates the secrets at the beginning of the protocol.  For the purpose of security, it should be possible for the issuer, after transferring its tokens to commercial banks, to be able to permanently delete all secrets related to the specific tokens that it issues.
    \item \textit{Modularity.} The overall architecture should rely upon simple components that can be rigorously analysed.
    \item \textit{Succinct audit log.} Well-known verifiable computation techniques (for example, proofs computed on Merkle trees) can be employed to make the audit log data succinct, so that it can be verified efficiently.
\end{itemize}

Finally, we believe that the design requirements can be met in two ways: either
(a) by having a minter sign blinded tokens and publish the signatures alongside
burned tokens on the audit log, as in the design proposed by Goodell, Toliver,
and Nakib~\cite{goodell2022}, and then having the payer furnish a
zero-knowledge proof that the unblinded token matches some blinded token on the
audit log; or (b) by having the payer's bank publish zero-knowledge commitments
to new tokens alongside burned tokens of equal value on the audit log and then
having the payer furnish a zero-knowledge proof that the new token manages some
zero-knowledge commitment on the audit log.  In this article, we shall formally
prove claims about the latter approach, although we assert that it is also
possible to prove similar claims about the former approach.

\section{Related Work}
Starting from the first work by Chaum~\cite{chaum1982}, digital cash has been the focus of much attention from industry and academia. \ignore{Especially in recent times, due to decentralized public ledgers with high integrity guarantees, such as blockchains, as well as the recent intention of many governments to implement central bank digital currencies, significant effort has been made to introduce new models for efficient digital currency systems with stronger security guarantees.

}At ACM CCS in 2022, two blockchain-aided digital currency systems were proposed: Platypus \cite{WustKDC22} and PEReDI \cite{KiayiasKS22}.  In terms of the definition of the privacy desiderata, we looked mainly to Platypus for inspiration.  In terms of efficiency and privacy guarantees, our design is comparable.  However, we recall that the Platypus protocol is account-based, and an important property that we require, transaction independence, cannot be satisfied by Platypus.  With account-based designs, users are required to hold long-term secrets that can be used to establish relationships among transactions.  Indeed, users who leak their own secrets in the Platypus system, whether by mistake, theft, or coercion, expose their entire account histories.

PEReDi is a robust payment protocol in which regulation can be strictly enforced by inspecting, in a privacy-preserving way, the payment traces of the users.  Its authors prove the security of their protocol in the Universal-Composability framework. However, like Platypus, PEReDi also uses an account-based design, which implies a persistent identity.  The PEReDi protocol is complicated, so Dogan and Bicakci responded by proposing KAIME \cite{DoganB24}, another account-based protocol with a simpler design, relying on a few simple cryptographic primitives.  However, although we recognise the value of the improvements, the account-based mechanism is still problematic because of the lack of transaction independence.

In AFT '20,  Androulaki et al. \cite{AndroulakiCCDET20} proposed a token-based system with auditability features.  Dogan and Bicakci considered the scheme of Androulaki et al. to be a ``unspent transaction output'' (UTxO) token scheme, because new tokens are created from the process of spending previous tokens.  The UTxO property holds whether or not the tokens themselves are written to the ledger, as they are in Bitcoin.  The design of Androulaki et al. uses zero-knowledge commitments to new tokens during a transaction, with the expectation that such tokens will be spent onward.  A UTxO architecture does not in itself imply a requirement for the payer to retain secrets from one payment to the next.  So it is possible to create a UTxO scheme that preserves transaction independence.  Nevertheless, the trust model of Androlaki et al. violates transaction independence for two reasons.  First, auditors are able to see and connect all of the transaction information, including all of the payments, that a payer makes, by design.  Second, authorisation for spending is bound to a user's identity, the credentials for which must be preserved across transactions.

Similarly, the trust model of Wüst et al.~\cite{WustKCC19}, another UTxO scheme, violates transaction independence for two reasons.  First, the payer must retain a private key corresponding to the certificate issued by the regulator enabling the payer to pay (that is, a long-lived identity).  It is assumed that the private key would be used by the payer to receive money when withdrawing it, for example, from a bank, and that the long-lived identity is linked to every transaction that the user makes as payer, even whilst it might not be disclosed.  Second, the payer must retain proof that previous transactions during an epoch were sufficiently small that the new payment does not violate a specified transaction limit.  These proofs ultimately add an account-like layer to what would otherwise be a UTxO token: A function of this transaction and previous transactions must be shown to meet a certain requirement.  Thus, we can argue that the architectures described in both the Androulaki et al. and Wüst et al. proposals are linked to an underlying identity of the payer.  Tomescu et al.~\cite{TomescuBAAGPY22} also proposed a token-based system in the UC setting.  As with PEReDi~\cite{KiayiasKS22}, we noticed a high design complexity, and as with the aforementioned UTxO protocols, there is an implicit requirement of a payer identity.
\ignore{
In our proposed architecture, a wallet does not carry any identifying information at all, and once money is paid, it is assumed to be gone, with nothing remaining that can be used to enforce limits or anything else.  Our architecture assumes that limits are enforced at the time at which consumers withdraw tokens, and not when they spend them, drawing into question an expectation anticipated by many designs that the identity of the payer must be linked to every transaction.  The mechanism we propose is sufficient to satisfy legitimate regulatory objectives, both because tokens once received must be exchanged via regulated intermediaries for their value to be spent anonymously in the future, and because it is assumed that regulators would require recipients of money to associate their own identities with the money they receive.
}

An effective alternative for privacy-preserving payments could be implemented by adapting the existing decentralized anonymous payment system such as ZeroCash~\cite{Ben-SassonCG0MTV14,zcash} and Monero~\cite{cryptonote,monero}. Their systems are based on zero-knowledge proofs and ring signatures, respectively, and are designed to provide privacy and integrity in a Bitcoin-like UTxO fashion for the fully decentralized context. Although such protocols would be easily adapted to institutional digital currency, such as CBDC, their usage could be considered overkill, given that a simpler approach to zero-knowledge proofs can be employed in our scenario since only the purchasers, and not the sellers, require anonymity. 
Indeed, although the bulletin board does not reveal any information about the seller, banks can link the seller's identity with the seller's tokens.
This highly facilitates the design and maintenance of the system.

Moreover, although our bulletin board could be implemented with a blockchain system, and our system does not forcibly require a blockchain system since any bulletin board with integrity features would be sufficient for our goals.

Our system decouples transactions wherein spent tokens are recycled into new tokens (burning) from spending transactions.  This implies that tokens can be spent without real-time access to the bulletin board and the recipient does not need to download the last state of the bulletin board. Indeed, it is sufficient for the payer to provide proof of inclusion and uniqueness (e.g., using a Merkle trie) of a signed update transferring a token to the recipient, together with a zero-knowledge proof certifying that the token had been correctly minted.

%
%

\section{Our Protocol}
\label{s:prot-overview}
The main idea behind the protocol is to allow a receiver of a token to spend it by posting a fresh and unrelated token onto a bulletin board without requiring any trusted party to verify the legitimacy of the new token.
Since this type of verifiability is unattainable without a trusted party or some relation between the tokens, we rely on non-interactive zero-knowledge proofs to create a relation between a freshly generated token and an older claimed token (which we refer to as {\em burnt}) without disclosing which was the exact token that was claimed in place of the new one.
With this approach, an eavesdropper looking at the bulletin board cannot identify which was the claimed token.
We call this kind of security against an eavesdropper  {\em token privacy}.

The main challenge behind this approach is in the generation of the the initial set of tokens, characterizing the overall token balance of the system, in a trusted manner. We call such tokens {\em genesis tokens}. For availability and security reasons, a central bank serves as issuer and shall be considered an available trusted party only during the protocol setup phase. After the setup phase, we do not have any trust assumption except that the bulletin board preserves availability and integrity of the posted data.
The recipients of genesis tokens are banks, who are a subset of the users and hence treated exactly as any other user transferring tokens inside the system.

To ensure that an adversary corrupting some subset of users cannot double spend one of his tokens, we introduce the concept of {\em validity} of a token.
A token is essentially composed of two keys, the {\em sender verification key}, freshly generated by the sender, and the {\em receiver verification key}, freshly generated by the receiver and sent to the sender to produce a new token. Together with such keys, other cryptographic objects such as signatures and zero-knowledge proofs might be part of the token to guarantee its validity.

A genesis token is valid if (i) the sender verification key is a central bank key and (ii) the signature attached to the token verifies the bank verification key w.r.t. the central bank verification key. These two checks are sufficient to identify valid genesis tokens inside the bulletin board subject to the trust assumption of the central bank and the verification keys the central bank has distributed during setup.
We further distinguish between {\em live} tokens and {\em burnt} tokens. A valid token is burnt if flagged as burnt by the receiver, and alive otherwise. The receiver further authenticates the burning procedure, i.e. the flag has a signature attached that verifies the token and the corresponding flag w.r.t. the receiver verification key.  Any token that is subsequent to a genesis token is considered a valid token when 
the token is {\em fresh}, meaning that the sender verification key was not used in an earlier valid token,
the token signature verifies the message composed of the receiver verification key with respect to the sender verification key,
the statement of the zero-knowledge proof, which is attached with the proof, refers to a set of burnt valid tokens, and the zero-knowledge proof is verifiable with respect to the attached statement.

To accept a new payment, the receiver verifies that the token posted by the sender onto the bulletin board is a live, valid token and that its associated verification key is correct, meaning that it matches the one that the receiver provided to the sender.

As it will be explained in more detail in the remainder of the paper, those validity checks ensure that the adversary cannot forge a new valid token without burning a live, valid token for which the adversary knows the signing key corresponding to the receiver verification key of the burnt token.  We refer to this property as {\em token integrity}.  We further require another property that the technique described above must satisfy, which we call {\em balance invariance}, wherein the overall number of live, valid tokens in the system (in the bulletin board), does not change its cardinality, i.e. it must always match the number of genesis tokens.

\label{s:our-protocol}


In the following, we first introduce some notation that will make the protocol easier to describe. Then, we introduce the NP-relation that must be proven by our NIZK-PoK system, and finally, we formally describe our protocol.
Because of space constraints, we deferred the definitions, the formal description of the protocol, and the proofs in \cref{s:preliminaries}, \cref{s:formal-desc} and \cref{s:proofs} respectively.

\paragraph{Auxiliary Notation and Useful Algorithms.}
Each time a new message is posted to a bulletin board $\bboard$, a local set of messages $\msgs$ is updated. Since $\bboard$, when receiving a new message, automatically notifies the parties with the new message, we assume that the set $\msgs$ of bulletin board messages is always updated by any involved party in the system.
We indicate as $\msgs[j]$ as the $j$-th token in $\msgs$.
Morever, we indicate as $\msgs_{live}$ as the set $\msgs$ restricted to live valid tokens and $\msgs_{burnt}$ as the $\msgs$ restricted to burnt valid tokens. Let $\indexes$ be a set of indexes, we indicate $\msgs^{\indexes\rceil}$ as the valid tokens in $\msgs$ restricted to the ones indexed by $\indexes$, i.e. the subset of $\msgs$ composed by the the tokens $(j,\cdot)$ for each $j\in\indexes$. When we are referring to the token with index $j$ in $\msgs$, we indicate it as $\restrict{M}{j}$.
We indicate a live token with sender $\sender$ and receiver $\receiver$ with $F^{\sender,\receiver}=(j,\vk^\sender,\vk^\receiver,\sigma^\sender,\indexes,\pi)$, where 
$j$ is the index of the token in the bulletin board,
$\vk^\sender$ is the fresh verification key of the sender,
$\vk^\receiver$ is the fresh verification key of the receiver,
$\sigma^\sender$ certifies that the sender owning $\vk^\sender$ authorizes the transfer to $\vk^\receiver$, i.e. $\sigma^\sender$  verifies the message $\vk^\receiver$ w.r.t. the key $\vk^\sender$,
the proof $\pi$ shows that $\vk^\sender$ is associated to one of the burning factors in the set of tokens $\restrict{\msgsburnt}{\indexes}$.
We indicate the genesis live token as $\token^{\sender,\receiver}=\{j,\vk^{\sender},\vk^{\receiver},\sigma^{\sender}\}$ with $\sender=\cbank$.

 A burnt token $\token^{\sender,\receiver}_{burnt}$ is additionally composed of a burning factor $\blinded^\receiver$ and the receiver signature $\sigma^\receiver$  certifying that the receiver acknowledged the token possession and burned it by using the burning factor $\blinded^\receiver$, i.e.
 $\sigma^\receiver$ verifies the message $(\vk^\sender, \blinded^\receiver)$ w.r.t. the key $\vk^\receiver$.
The burning factor $\blinded^\receiver$ is a committed freshly generated sender verification key that will be used by $\receiver$ to spend such token.

To extract valid burnt tokens $\msgs_{burnt}$ from $\msgs_{valid}$ the user run an algorithm $\getburnt$ outputting all the tokens in $\msgs_{valid}$ in which the sender verification key and the burning factor $\blinded$ associated with the token are certified by the receiver, i.e. such that the signature $\sigma^\receiver$ attached to the burnt token verifies the message $(\vk^\sender,\blinded)$ w.r.t. the key $\vk^\receiver$.

As an auxiliary useful algorithm, we use a validity predicate $\validity$. The algorithm  $\validity$ takes as input the set of central bank verification keys $VK$ valid tokens $\verassets$, a token $\token$ as above, computes $\msgs_{burnt}=\getburnt(\verassets)$ and, if $F$ is a standard token,  outputs $1$ if
there is no token in $\msgsburnt$ with the same sender verification key,
$\sigma^\sender$ verifies the message $\vk^\receiver$ w.r.t. the key $\vk^\sender$,
the size of $\restrict{\msgsburnt}{\indexes}$ is the same of $\indexes$ and the proof $\pi$ verifies w.r.t. the burning factors in $\restrict{\msgsburnt}{\indexes}$.

Otherwise, if $F$ is a genesis token, output $1$ if $\vk^\sender\in VK$ and $\sigma^\sender$ verifies the message $\vk^\receiver$ w.r.t. the key $\vk^\sender$.

As we will see in the formal description, this algorithm will be used by each user when a new token appears on the bulletin board to locally construct the set of valid tokens $\msgs_{valid}$.

\paragraph{Our NIZK-PoK Relation.}
Let $\Commit$ be a commitment scheme (\cref{s:com}). Our non-interactive zero-knowledge proof of knowledge (NIZK-PoK) shows that the sender verification key of the new token is associated to a burnt valid token inside a bucket of burning factors. 
Burning factor are, in practice, committed verification keys.
Let us take the $n$ burning factors $\blinded_1,\ldots,\blinded_n$. The relation must show that the sender verification key $\vk$ freshly generated by the receiver of a burnt token is committed in one of the burning factors, i.e. $\Commit(\vk;r)=\blinded_i$ for some $i$ and some randomness $r$, without exposing in which burning factor was committed, i.e. the index $i$ is hidden. Hence, the randomness $r$ and the index $i$ must be part of the witness.
Formally, the relation to be proven is:

\begin{equation}
\Reldc^\Commit=\{((\blinded_1,\ldots,\blinded_n,\vk),(r,i)) : \Commit(\vk;r)=\blinded_i\}.
\end{equation}

\paragraph{On the efficiency of the proof generation.} Notice that a proof for the relation proposed is fairly simple to implement in an efficient manner. In fact, this relation can be cast as a $1$-out-of-n Proof-of-Partial Knowledge (PPK), where the prover proves knowledge of a witness out of $n$ statements without revealing for which of the $n$ statements the prover knows the witness. Indeed, the statement above can be re-casted as a $1$-out-of-$n$ proof where $\stm_i=(\blinded_i,\vk)$ and $\wit=(r,i)$.
As shown by \cite{GoelGHK22}, given a $\Sigma$-protocol for a certain language, it is possible to construct a communication-efficient $1$-out-of-$n$ PPK where each $\stm_i$ is an instance of the base $\Sigma$-protocol. The resulting PPK has size $O(\log n)$ if the underlying $\Sigma$ protocol has constant size proofs.
By instantiating $\Commit$ with a Pedersen commitment \cite{Pedersen91}, it would be sufficient to use a $\Sigma$-protocol for Discrete Log equality to check $\Commit(\vk;r)=\blinded_i$.
Between the possible improvements of our protocol, we might make the honest user to spend $t$ tokens all at once and provide a cumulative NIZK-PoK showing that $t$ previous tokens have been previously burned.
This can be done via the $t$-out-of-$n$ PPKs, ensuring that the prover knows the witness of at least $t$ out of $n$ statements.
The size of the $1$-out-of-$k$ PPK of \cite{AttemaCF21} for DL-like languages (such as DL equality) is only of $O(\log n)$.

We notice that even general-purpose zk-SNARKs \cite{Groth16,BitanskyCIPO13a}, usually requiring prohibitive computational costs for large circuits, would not be totally impractical for the circuit describing $\Reldc^\Commit$, especially when using ZK-proofs friendly commitments such as the Pedersen commitment. The proof size of zk-SNARKs would be constant.

We remark that even though the strongest version of our protocol requires all the parties to monitor the bulletin board to verify updates periodically, this task can be delegated to the banks.


\paragraph{Protocol Description.}

The formal description of our protocol $\pidc$ is given in \cref{s:formal-desc}. The protocol is initialized with a central bank $\cbank$ and a bulletin board $\bboard$ formally described in \cref{fig:funccb} and \cref{fig:funcbb} in \cref{s:formal-desc} respectively, a set of users $\users$ and a subset $\banks\subseteq\users$ of banks.

The protocol is divided into 3 main phases: Setup, Burning, and Spending. In the setup phase, which is run only once at the beginning, genesis tokens, composed of a freshly generated verification key of the central bank $\vkcb^{(i,\bank)}$ and a freshly generated verification key $\vk^i_{\bank}$ from a bank $\bank\in\users$, are created by the central bank $\cbank$ and posted to $\bboard$ in a signed fashion (the signature $\sigma^{(\bank,i)}_{\cbank}$ verifies the message $M=\vk^i_{\bank}$ verifies under the key $\vkcb^{(i,\bank)}$). Notice that we assume that each central bank's verification key $\vk^{(i,\bank)}$ is trusted and certified by the central bank.

As soon as the genesis tokens are added into $\bboard$, each user $\user$ (which are banks in the case of genesis token) verifies their validity and adds them into their validity set $\msgs_{valid}^\user$. Since we are considering genesis tokens, which are trusted by assumption, it is sufficient that the validity algorithm checks their signature $\sigma^{(\bank,i)}_{\cbank}$ against $M=(\vk^i_\bank)$ and key $\vkcb^{(i,\bank)}$.

When an honest user $\receiver_0$ receives a token, they immediately burn it. Notice that each time a new token, live or burnt, is posted into $\bboard$, each honest user $\user$ verifies its validity and adds it to its local set of valid tokens $\msgs_{valid}^\user$ (auxiliary Token Validation phase).
Hence, if the new live token is part of the receiver's validity set $\msgs_{valid}^{\receiver_0}$, $\receiver_0$ can safely burn and lately spend it.

In the burning phase, $\receiver_0$, which will become a future sender $\sender_1$, first creates a fresh verification key $\vkso$ (Create sub-phase). Then augments its live token with a burning factor (commitment) $\blindedso$, embedding the freshly generated verification key $\vkso$, together with a signature $\sigma^{\receiver_0}$ certifying its authorization to burn such token (Burn sub-phase). This signature verifies the message $M=(\vksz,\blindedso)$, where $\vksz$ is the verification key of the sender of such burnt token, under the verification key $\vkrz$ that the receiver used for that token.

When new sender $\sender_1$ (formerly $\receiver_0$) decides to spend the token after interacting with the future receiver $\receiver_1$, it first waits to receive a verification key $\vkro$ from $\receiver_1$ (Token Gen sub-phase). Then, $\sender_1$ generates a NIZK-PoK showing that it owns one of the verification keys embedded into one of the elements of a randomly chosen set of burning factors inside the burnt tokens set (Proof Gen sub-phase).
During proof generation, $\sender_0$ further generates a signature $\sigso$ certifying that they can spend such token with the receiver verification key $\vkro$ (the signature verifies $M=\vkro$ under the verification key $\vkso$).
Finally, $\sender_1$ through their bank, which performs some regulation check, posts the live token into $\bboard$ (Token Post sub-phase).

To recap, a live token $\token$ is composed of a sender verification key $\vkso$, a receiver verification key $\vkro$, a signature $\sigma^{\receiver_1}$ and a proof $\pi$ with a set of indexes $\indexes$ connecting such proof with a set of burning factors together the sender verification key $\vkso$.

The knowledge soundness property of $\pi$, in conjunction with the validity requirement that no sender verification key can be re-used across the tokens (checked by the honest users during the Token Validation phase through the algorithm $\validity$) and by the computational binding property of the committed verification keys (burning factors), ensures that no adversary can create a new token from scratch or double-spend an already owned one.
The zero-knowledge property of $\pi$, coupled with the hiding property of the committed verification keys, avoids the disclosure of the burning factors embedding the sender verification key of the token, ensuring privacy.
Formal details are provided in the next section.

\paragraph{Efficiency Improvements.}
In terms of communication complexity, notice that each user has to download each token posted on the bulletin board and keep a set of indices related to valid tokens. In practice, each user has to download $O(\#\text{tokens})$ tokens and keep them in storage $O(\#\text{tokens})$ indexes to keep track of the validity of such tokens.

Other than the communication burden, we notice a computational burden. Indeed, each player has to verify a NIZK-PoK each time a new token is posted, and when spending, a NIZK-PoK proof must be generated.
However, an external party might take all the NIZK-PoKs posted until that moment and produce a succinct aggregated proof that can be verified faster than verifying every single proof (for aggregation, see, for example, \cite{GaillyMN22}). This incurs both communication and computational savings for each user.

Regarding the problem of generating a NIZK-PoK each time a token must be spent, we notice that several privacy-preserving proof outsourcing mechanisms have been developed, e.g. \cite{GargG0PS23}.

Regarding the potentially huge size of the token set in the bulletin board, we notice that it is possible to use common verifiable computation techniques, such as recursive snarks on hashed data, to make the user download just a succinct representation of data together with a proof certifying its correctness. This idea has been adopted in the context of blockchain protocols by MINA \cite{MINA}.

Besides our protocol being inspired by \cite{goodell2022}, our technique diverges from that work due to our use of ZK proofs to guarantee stronger security properties.
Rather than initialising the blind signature scheme, the consumer first requests the identity of a spent token, which the consumer then uses to create a burning factor (a commitment $\beta$) to a new token and a
signature linking that commitment to the spent token ($F$).  The consumer
sends $F$ and $\beta$ to the bank, which writes them to the bulletin board and
provides evidence to the consumer that this is done.
The consumer optionally sends $\vkro$ and $\sigso$
to the merchant, which enables the merchant to verify the validity of the token
in ZK (by looking at the bulletin board), avoiding the risk that the minting plate will have been consumed
following the creation of invalid tokens by minters.

\section{Security Analysis}
\label{s:analysis}

In the following, we introduce the main security properties we require and show that they are satisfied by the system.
\paragraph{Token integrity.} We divide the token integrity requirement into two main properties, {\em token forgery} and {\em balance invariance}.
\begin{definition}[Token Forgery]
Given our system and users $\users$,
the game consists of an interaction between an adversary $\adv$  and
a challenger $\chal$ with access to an oracle $\oracle$ that simulates honest
parties in the system. 
The game proceeds as follows:
    \begin{enumerate}
        \item $\chal$ initializes the system with the security parameter $\secpar$, instantiates all the necessary configuration values, such as the CRS of the NIZK proof of knowledge system and
      initializes the oracle $\oracle$
        creating tokens and controlling honest parties.

        \item $\adv$ communicates to the challenger a subset of users $\corr\subset\users$ he wishes to corrupt and tells $\oracle$ how many tokens must be assigned to each user in $\users$.
        \item $\oracle$ sends to $\adv$ the secrets related to the tokens meant for users in $\corr$.
        \item $\adv$ can receive tokens from the parties controlled by $\oracle$ and transfer tokens\footnote{We recall that, in our protocol,  transferring is done by creating a fresh token and burning and old one in a verifiable and private fashion.} to parties controlled by $\oracle$. Furthermore, $\adv$ can ask the oracle to transfer tokens between parties controlled by $\oracle$. All token transferred from an interaction with $\chal$ are inserted into a query set $\calQ$.
        \item $\adv$ wins the game if he can create a token not existing in $\calQ$ accepted by $\oracle$,
        in which $\adv$ does not control the sender nor recipient, or an asset in which $\adv$ controls the recipient,
but not the sender and no asset with the
 and the same value exists in $\calQ$.
    \end{enumerate}
\end{definition}
\begin{Claim}[Token Unforgeability]\label{claim:tokenunf}
    No PPT $\adv$ without access to the  extraction trapdoor of the zero-
knowledge proof system can win the token forgery game with
non-negligible probability w.r.t. protocol $\pidc$.
(For proof of this claim, refer to \cref{app:tokenunf-proof}.)
\end{Claim}

\begin{Claim}[Balance Invariance]\label{claim:balinv}
No PPT $\adv$ without access to the simulation trapdoor of the zero-knowledge
the proof system can create an asset
that increases the available
funds in the system, i.e., each freshly created token must be bound to exactly one burnt token.
\end{Claim}
\begin{proof}
    Balance invariance follows straightforwardly from token unforgeability. Indeed, no adversary is eable to modify the overall balance of the system, initially fixed by the central bank through the distribution of the genesis tokens.
\end{proof}
\paragraph{Token Privacy.} To ensure token privacy, we require a token indistinguishability property, defined below.
\begin{definition}[Token Indistinguishability]
    Given our system and users $\users$,
the game consists of an interaction between an adversary $\adv$  and
a challenger $\chal$ with access to an oracle $\oracle$ that simulates honest
parties in the system.
The game proceeds as follows:
\begin{enumerate}
   \item $\chal$ initializes the system with the security parameter $\secpar$.
        $\chal$ instantiates all used primitives, such as the signature schemes or the NIZK proof-of-knowledge system. $\chal$ further initializes the oracle $\oracle$
        creating assets and managing token exchange with and between honest parties.
            \item $\adv$ communicates to the challenger a subset of users $\corr\subset\users$ he wishes to corrupt and tells $\oracle$ how many tokens must be assigned to each user in $\users$.
        \item $\oracle$ sends to $\adv$ the secrets related to the tokens meant for users in $\corr$.
             $\adv$ can ask to receive tokens from the parties controlled by $\oracle$  tokens\footnote{We recall that, in our protocol,  transferring is done by creating a fresh token and burning and old one in a verifiable and private fashion.} to parties controlled by $\oracle$ (using their secrets). Furthermore, $\adv$ can ask the oracle to transfer tokens between honest parties controlled by $\oracle$ using two oracles $\oracle_1$ and $\oracle_2$. The adversary shall produce, through interaction with such oracles, two different valid payment traces of the same size where only token transfers between honest players can differ.

             This operation is done statically, i.e. the adversary determines all the token transfers in a single query, i.e. a single shot interaction with $\oracle,\oracle_1$ and $\oracle_2$.
        \item $\chal$ chooses a bit $b\getsr\bin$  and sends the payment trace (the set of token transfers) produced by $\oracle$ and $\oracle_b$ to $\adv$.
        \item $\adv$ outputs a bit $b'$ and wins the game if $b=b'$.
\end{enumerate}
\end{definition}
\begin{Claim}[Token Indistinguishability]\label{claim:tokenind}
     No PPT $\adv$ without access to the trapdoor can win the token indistinguishability game w.r.t. protocol $\pidc$ with
non-negligible probability.
    (For proof of this claim, refer to \cref{app:tokenind-proof}.)
\end{Claim}

\paragraph{Transaction Independence.} We further introduce another property, which we call {\em transaction independence}, stating that, if an adversary corrupts an honest user $\user$ during the system execution, cannot learn more than the information corresponding to received tokens the user has not spent.
This implies that the payment history of the user is kept secret even if the user is corrupted.

\begin{Claim}\label{claim:transind}
   No PPT adversary $\adv$ without knowledge of the trapdoor can break transaction independence of $\pidc$.
\end{Claim}
\begin{proof}
    In our protocol $\pidc$, transaction independence is satisfied since each user, after burning a received token, keeps in their storage the signing key of the corresponding receiver verification key $\receiver_0$, the commitment opening values $(\vk^{\sender_1},r)$ and the secret  $\sk^{\sender_1}$.
After spending the token, the user does not need to hold such values, which can be safely eliminated from their local storage.

\end{proof}

\begin{theorem}
    Let $\SIG$ be a Signature Scheme and $\nizk$ be a NIZK-PoK for the relation $\Rel$ defined above.
    Then, $\pidc$ satisfies transaction unforgeability, balance invariance, transaction privacy and transaction independence.
    (The proof follows by combining \cref{claim:tokenunf} to \cref{claim:transind}.)
\end{theorem}

\section{Conclusions and future directions}
\label{s:conclusion}
We proposed a novel protocol allowing private electronic payments with self-custody and zero-knowledge verified assurance. Our protocol ensures consumer privacy, compatibility with regulatory objectives, independent transactions in the sense that secrets related to a specific transaction do not leak any information about the sender or receiver transaction history, and no involvement of the issuer in the circulation of tokens. The issuer (central bank) is further stateless in the sense that after the issuer distributes the public values connected with the secret it had locally generated, the issuer can eliminate such secrets and disappear from the equation.

In terms of efficiency guarantees, re-issuance of tokens can be performed in an efficient manner, as well as for proof generation.
The bulletin board size, as well as the communication complexity and the storage consumption of the users, might be prohibitive with the system growing. As pointed out, proof aggregation and verifiable computation techniques, coupled with the aid of external parties whose task is to produce computationally intensive recursive proofs, could help solve the issue.

As future work, we aim to provide a second protocol with weaker but reasonable security guarantees about the validity of assets, with enhanced efficiency that maintains a simple design and relies on simple cryptographic objects, such as blind signatures, but relaxing the requirement for all end-user devices to produce and verify zero-knowledge proofs.  We envision that this protocol would be useful for some payment scenarios for which ZKP verification would be impractical.
In particular, we would rely on an external semi-trusted entity, a minter, taking the burden of minting new tokens. Since such minters might be corrupted, we would employ a fail-safe mechanism based on a mixture of zero-knowledge proofs and a law-enforced auditing mechanism to pinpoint and kick out dishonest miners from the system. This would be accompanied by a mechanism accounting for eventual currency loss due to the minter's misbehaviour.  We discuss this protocol idea in~\cref{s:alternative}.

\section*{Acknowledgements}

The authors acknowledge project SERICS (PE00000014) under the MUR National
Recovery and Resilience Plan funded by the European Union -- NextGenerationEU.
The authors would like to thank Professor Tomaso Aste for his continued support
of this work, as well as Ching Yeung Fung for providing feasibility testing.
The authors also acknowledge the UCL Future of Money Initiative and the
Systemic Risk Centre at the London School of Economics.

\bibliographystyle{plain}
\bibliography{biblio}


\appendix

\section{Useful Tools}

\label{s:preliminaries}
\subsection{Signatures}
A signature scheme is a tuple of algorithms $\Pi = (\kgen, \sign, \ver)$ with message space $\msgspace$, described as follows:
\begin{description}
    \item[$\kgen(1^\secpar)$:] On input the security parameter, outputs a verification key/signing key pair $(\vk, \sk)$.
    \item[$\sign(\sk, \msg)$:] On input a signing key $\sk$ and a message $\msg \in \msgspace$, outputs a signature $\sigma$.
    \item[$\ver(\vk, \msg, \sigma)$:] On input a verification key $\vk$ and a message $\msg \in \msgspace$, outputs $1$ if the signature verifies, and $0$ otherwise.
\end{description}

\begin{definition}[Correctness]
A signature scheme $\Pi$ is correct if, for all $\msg \in \msgspace$,
$\prob{\ver(\vk, \msg, \sign(\sk, \msg)) = 1}$,
where $(\vk, \sk) \getsr \kgen(1^\secpar)$
\end{definition}
\begin{definition}[EUF-CMA]
    A signature scheme $\Pi$ is existentially unforgeable under chosen message attacks if, for all valid  PPT  $\adv$,
 \[
 \prob{
 \begin{tabular}{c}
 $\ver(\vk,\msg, \sigma)=1 \land$\\
 $\msg \text{ is fresh}$
 \end{tabular}
    \middle |
        \begin{tabular}{c}
            $(\vk, \sk) \getsr \kgen(1^\secpar)$;  \\
            $(\msg,\sigma)\getsr\adv^{\sign(\sk,\cdot)}(\vk)$
        \end{tabular}
    } \le\negl.
    \]
\end{definition}
By ``fresh'', we mean that it was never queried to the $\sign(\sk,\cdot)$ oracle by $\adv$.
\subsection{Blind Signatures}
A blind signature is composed of a tuple of algorithms $\BS=(\Gen,\sig,\bsign,\allowbreak\blind,\unblind,\ver)$ described as follows:
\begin{description}
\item[$\Gen(1^\secpar)$:] The key generation algorithm take as input the security parameter $1^\secpar$, and outputs a pair of signing and verificaiton key $(\vk,\sk)$.
\item[$\blind(\msg)$:] The blinding algorithm takes as input a message $\msg$ and outputs a blinded message $\blinded$ together with its blinding factor $r$.
\item[$\bsign(\sk,\blinded)$:] The signature algorithm takes as input a blinded message $\blinded$, a signing key $\sk$, and outputs a blind signature $\bsigma$.
\item[$\sig(\sk,\blinded)$:] The signature algorithm takes as input a message $\msg$, a signing key $\sk$, and outputs a signature $\sigma$.
\item[$\unblind(\blinded,\bsigma,r)$:] The unblinding algorithm, on input a blinded message, a signature on the blinded message $\bsigma$, and a blinding factor $r$, outputs the signature $\sigma$ on the unblinded message $\msg$.
\item[$\ver(\vk,\msg,\sigma):$] The verification algorithm outputs $1$ if $\sigma$ is a correct signature of $\msg$ upon verification key $\vk$, 0 otherwise.
\end{description}
\begin{definition}[Correctness] A signature scheme $\BS$ is correct  if
$$
\prob{\ver(\vk,\msg,\sigma)=1 \ \middle | \ 
\begin{tabular}{c}
$(\vk,\sk)\getsr\Gen(1^\secpar);$ \\
$(\blinded,\bfactor)\getsr\blind(\msg)$,\\ 
$\bsigma\getsr\bsign(\sk,\blinded)$;\\
$\sigma=\unblind(\tsigma,\bfactor)$
\end{tabular}
}=1$$
\end{definition}
Security of Blind Signature is defined w.r. Existential Unforgeability w.r.t Blinded Messages and Blindness
\begin{definition}[Existentital Unforgeability w.r.t Blinded Messages]
A blind signature scheme $\BS$ is Existentially Unforgeable w.r.t Blinded Messages, if for all PPT $\adv$,
$$\prob{\game^{\bEUFCMA}_{\BS,\Adv}(\secpar)} \le \negl,$$
where the challenger of $\game^{\bEUFCMA}_{\BS,\Adv}(\secpar)$ behave as follows:
\begin{description}
   \item[Setup phase:] Generate $(\vk^*,\sk^*)\gets\Gen(1^\secpar)$. Then, initialize a query counter $\sfQuery=0$.
   \item [Query phase:] When receiving a blinded value $\blinded$ from $\adv$, invoke $\tsigma\getsr\bsign(\sk^*,\blinded)$ and update $\sfQuery\gets\sfQuery+1$. Then, send $\sigma$ to $\adv$.
   \item[Challenge phase:] When receiving $\{(\msg_i,\sigma_i)\}_{i\in[\sfQuery+1]}$ from $\adv$, check that $\ver(\vk^*,\msg_i,\sigma_i)=1$ for each $i\in[\sfQuery+1]$ and that $\msg_i\ne\msg_j$ for each $i,j$ with $i\ne j$. If the checks pass, 
  output 1. Else, output 0.
\end{description}

\end{definition}
\begin{definition}[Blindness] 
A blind signature scheme $\BS$ enjoys blindness if, for all PPT adversary $\adv=(\adv_0,\adv_1)$,
$$
\prob{b=b^* \ \middle | \ 
\begin{tabular}{c}
$(\msg_0,\msg_1)\getsr\adv_0(1^\secpar)$;\\
$b^*\getsr\bin$; \\
$(\blinded,\bfactor)\getsr\blind(\msg_b)$;\\ 
$b\getsr\adv_1(\msg_0,\msg_1,\blinded)$.
\end{tabular}
}\le\frac{1}{2}+\negl.$$
\end{definition}

Throughout this paper, we use the abbreviation PPT to denote probabilistic polynomial time. 
Given a PPT algorithm $\adv$, let $\adv(x)$ be the probability distribution of the output of $\adv$ when run with $x$ as input.
We use $\adv(x;r)$, instead, to denote the output of $\adv$ when run on input $x$ and coin tosses $r$.
We denote with $\secpar\in\NN$ the security parameter and with $\Poly(\cdot])$ an arbitrary positive polynomial. 
Every algorithm takes as input the security parameter $\secpar$ (in unary, i.e.\ $1^\secpar$). When an algorithm takes more than one input, $1^\secpar$ is omitted.
We say that a function $\nu: \NN\rightarrow\RR$ is negligible in the security parameter $\secpar\in\NN$ if it vanishes faster than the inverse of any polynomial in $\secpar$, i.e. $\nu(\secpar)<\frac{1}{\poly}$ for all positive polynomials $\poly$.

We use $\gets$ when the variable on the left side is assigned with the output value of the algorithm on the right side.
Similarly, when using $\getsr$, we mean that the variable on the left side is assigned a value sampled randomly according to the distribution on the right side.

A {\em polynomial-time} relation $\Rel$ is a relation 
for which membership of $(\stm,\wit)$ in $\Rel$ can be decided in time polynomial in $|\stm|$. 
If $(\stm,\wit)\in\Rel$, then we say that $\wit$ is a {\em witness} for the {\em instance} $\stm$.
A polynomial-time relation $\Rel$ is naturally associated with 

A distribution ensemble $\{X(\secpar)\}_{\secpar\in\NN}$ is an infinite sequence of probability distributions, where a distribution $X(\secpar)$ is associated with each value of $\secpar\in\NN$. 
We say that two distribution ensembles $\{X(\secpar)\}_{\secpar\in\NN}$ and $\{Y(\secpar)\}_{\secpar\in\NN}$ are {\em computationally indistinguishable} if for every PPT distinguisher $\advD$, there exists a negligible function $\nu$ such that:
\[
\prob{\advD(1^\secpar,X(\secpar))=1}-\prob{\advD(1^\secpar,Y(\secpar))=1} \leq \nu(\secpar).
\]
$\{X(\secpar)\}_{\secpar\in\NN}$ and $\{Y(\secpar)\}_{\secpar\in\NN}$ are statistically indistinguishable if the above holds for computationally unbounded $\advD$. We use $\approx_c$ and $\approx_s$ to denote that two distributions ensembles are respectively computationally and statistically indistinguishable,
$\equiv$ to denote that two distribution ensembles are identical.

\subsection{Non-Interactive Commitments.}\label{s:com}
A non-interactive commitment is a PPT algorithm $\Commit$ taking as input a message $\msg\in\bin^k$ and outputting a value $\com = \Commit(\msg;r)\in\bin^l$ where $r\in\bin^\secpar$ is the randomness used to generate the commitment. The pair $(\msg,r)$ is also called the {\em opening}. 
A non-interactive commitment typically satisfies two properties known as binding and hiding; we review these properties (in the flavor we need them) below.

\begin{definition}[Computational binding]
We say that $\Commit$ satisfies computational binding if, for all PPT adversary $\adv$, it holds that
\[
\prob{ \Commit(\msg_0;r_0) = \Commit(\msg_1,\allowbreak;r_1) \ \big | \ (\msg_0,\msg_1,r_0,r_1)\getsr\adv(1^\secpar) }\le\negl
\]
\end{definition}

\begin{definition}[Statistical hiding]
We say that $\Commit$ satisfies statistical hiding if for all pairs of message $\msg_0,\msg_1\in\bin^k$, it holds that $\{\Commit(1^\secpar,\msg_0)\}_{\secpar\in\NN} \approx_s\{\Commit(1^\secpar,\msg_1)\}_{\secpar\in\NN}.$
\end{definition}

\subsection{Non-Interactive Zero-Knowledge Proof-of-Knowledge.}

In the following, we introduce our NIZK-PoK definition\footnote{Since knowledge soundness is defined with a PPT adversary, it should be referred as be Argument-of-Knowledge. Throughout this paper, we keep using the term Proof in place of Argument.}.
A non-interactive proof system $\nizk$ is composed of a tuple of algorithms described as follows:
\begin{description}
	\item [$\setup(\Rel)$:] On input the NP-relation $\Rel$ returns a common reference string $\crs$ and a  trapdoor $\td$.
	\item [$\prove(\crs,\stm,\wit)$:] Take as  input the $\crs$, a statement $\stm$ and a witness $\wit$ such that $(\stm,\wit)\in\rel$ and returns a proof $\pi$.
	\item [$\ver(\crs,\stm,\pi)$:] Takes as input the $\crs$, a statement $\stm$ and a proof $\pi$. Output $1$ if $\pi$ is an accepting proof with respect to the statement $\stm$ and relation $\rel$, 0 otherwise.
 \end{description}

\begin{definition}[Non-Interactive Zero-knowledge  Proof-of-Knowledge] Let $\rel$ be an NP-relation. A non-interactive proof system $\nizk=(\setup,\prove,\allowbreak\ver)$ is a NIZK-PoK for the relation $\Rel$ if it satisfies completeness, knowledge soundness and statistical zero-knowledge. The properties follow.
\begin{itemize}
    \item \textbf{Completeness} For each  $(\stm,\wit)\in\rel$, we have that
$$ \prob{\ver(\stm,\wit,\pi)=1
\bigg | 
\begin{tabular}{c}
$(\crs,\td)\getsr\setup(\Rel);$ \\
$\pi\getsr\prove(\crs,x,w)$
\end{tabular}
}=1 $$
\item \textbf{Knowledge Soundness} For all PPT $\adv$ there exists an extractor $\ext_\adv$  such that, 
    $$  \prob{
    \begin{tabular}{c}
        $\ver(\stm,\pi)=1 \ \land$ \\
         $(\stm,\wit)\notin\rel$     
    \end{tabular}
  \ \bigg | \ 
\begin{tabular}{c}
$(\crs,\td)\getsr\setup(\Rel);$\\
$((\stm,\pi);\wit)\getsr\adv||\ext_\adv(\crs)$
\end{tabular}
    } \le \negl.$$

    \item \textbf{Statistical Zero-Knowledge}
There exists a simulator $\Sim$ such that for all $(\stm,\wit)\in\Rel$ and for all PPT  $\Adv$,
\begin{multline*}
    \Big\{\pi\getsr\prove(\crs,\stm,\wit) : (\crs,\td)\getsr\setup(\Rel)\Big\}\approx_s \\\Big\{\pi\getsr\Sim(\crs,\td,\stm) : (\crs,\td)\getsr\setup(\Rel)\Big\}
\end{multline*}


\end{itemize}
\end{definition}

Since, for communication efficiency reasons, we aim for short proofs, we further introduce the {\em succintness} property. Succint NIZK-PoKs are called zero-knowledge Succint Non-Interactive Arguments-of-Knowledge (zk-SNARKs).
\begin{definition}[zk-SNARKs]
    Let $\nizk$ be a NIZK-PoK for a relation $\Rel$. 
    We say that $\nizk$ is a zk-SNARK if it further satisfies succintness, gureanteeing that for any $(\stm,\wit)\in\Rel$, given $\pi$ be the output of $\prove(\stm,\wit)$, we have that $|\pi|=\poly$ and the verifier, on input $\stm$ and $\pi$, runs in 
    $\Poly(\secpar+|\stm|)$.
\end{definition}

 \section{Protocol Details}\label{s:formal-desc}

  The protocol is initialized with a central bank $\cbank$, a bulletin board $\bboard$ formally described in \cref{fig:funccb} and \cref{fig:funcbb}, a set of users $\users$ and a subset $\banks\subseteq\users$ of banks. 
 Let $\Commit$ be a commitment scheme with message space $\msgspace=\bin^\psi$, and commitment space $\comspace=\bin^\ell$ for some  $\ell, \psi \in\poly$. Let $\SIG=(\Gen,\sign,\ver)$ be a signature scheme with message space $\msgspace=\bin^*$, signature space $\sigspace=\bin^\kappa$ and verification key space $\vkspace=\bin^\psi$ for some $\kappa \in \poly$ and $\nizk=(\setup,\prove,\ver)$ be a NIZK-PoK for the relation $\Reldc^{\Commit}$.
The protocol is composed of the following phases:
\begin{description}
\item[Setup:] Each bank $\bank$ performs the following steps:
\begin{itemize}
    \item Compute $(\vk^i_\bank,\sk^i_\bank)\gets\SIG.\Gen(1^\secpar)$ for each $i\in[\tau]$ and sends $\{\vk^i_\bank\}_{i\in[\tau]}$ to $\cbank$.
\end{itemize}
Then, each user in $\users$ performs the following steps:
\begin{itemize}
\item Wait to receive $VK_{\cbank}=\{\vkcb^{(j,\bank)}\}_{j\in[\tau],\bank\in\banks}$ from $\cbank$.
    \item Fetch $(\crs,\sigma^{\crs}_{\cbank})$ from $\msgs$ such that $\SIG.\ver(\vk^\bank,\crs,\sigma^{\crs}_{\cbank})$.
    \item Initialize a set of valid tokens $\verassets\gets\emptyset$.
    \item For each $j\in[|\msgs|]$, if $\validity(VK_{\cbank},\msgs[j],\verassets^\user)=1$, set $\verassets^\user\gets\verassets^\user\cup\{\msgs[j]\}$.

\end{itemize}
    \item[Burning:] 
    A sender $\sender_1=\receiver_0\in\users$, the receiver of a token\footnote{If it is a genesis token, its form would be $\token^{\sender_0,\receiver_0}=\{j,\vksz,\vkrz,\sigsz\}$.} $F^{\sender_0,\receiver_0}=(\vksz,\vkrz,\allowbreak\sigsz,\indexes,\pi^0)\in\msgs_{valid}^{\sender_1}$ interact with their bank $\bankso$ and $\bboard$  as follows:
    \begin{enumerate}
        \item[{\it Create:}] Compute $(\vkso,\skso)\getsr\SIG.\Gen(1^\secpar)$.
     \item[{\it Burn:}] Compute $\blindedso=\commit(\vkso;r)$ for $r\getsr\bin^\secpar$,  $\sigma^{\receiver_0}\getsr\SIG.\sign(\skrz,\allowbreak(\vksz,\blindedso))$ and sends $\token^{\sender_0,\receiver_0}_{burnt}=(\token^{\sender_0,\receiver_0},(\sigma^{\receiver_0},\blindedso))$ to $\bankso$, forwarding, in turn,  $(\post,\token^{\sender_0,\receiver_0}_{burnt})$ to $BB$.
    \end{enumerate}
 \item[Spending:] 
A sender $\sender_1$, on input the burnt token $F^{\sender_0,\receiver_0}_{burnt}$,  the burning randomness $r$ and the new fresh key pair $(\vkso,\skso)$ interact with the receiver $\receiver_1$ as follows:
 \begin{itemize}
    \item[{\it Token Gen:}] $\receiver_1$ computes $(\vkro,\skro)\getsr\SIG.\Gen(1^\secpar)$ and sends $\vkro$ to $\sender_1$.
     \item[{\it Proof Gen:}] $\sender_1$ proceeds as follows:
     \begin{enumerate}
         \item Compute a signature $\sigso\getsr\SIG.\sign(\allowbreak\skso,\allowbreak\vkro)$ and choose a random set of indexes $\indexes$ of cardinality $n$ between the indexes of the tokens in $\msgs_{valid}$, including the index $k$ of the token burned using $\blindedso$.
         \item  Compute $\pi\getsr\nizk.\prove((\crs,\blinded_1,\ldots,\blinded_n,\vkso),(r,k))$ where $\blinded_i$ is the burning factor of $\restrict{\msgs_{burnt}}{\indexes}[i]$. Finally, send $(\sigso,\pi,\indexes)$ to $\receiver_1$, which sends $\token^{\sender_1,\receiver_1}$=  $(\vkso,\vkro, \sigso,\pi,\indexes)$ to $\bankso$.
     \end{enumerate}

     \item[{\it Token Post:}] $\bankso$, on input $\token^{\sender_1,\receiver_1}$, if $\checkreg(\token^{\sender_1,\receiver_1},\msgs)=0$ (i.e., the asset is not regulation compliant) send $(\denied,\token^{\sender_1,\receiver_1})$ to $\sender_1$, who forwards it to $\receiver_1$.
      Else, send $(\post,\token^{\sender_1,\receiver_1})$ to $\bboard$.
     \item[{\it Validation:}] $\receiver_1$, when receiving  $(\stored,\token^{\sender_1,\receiver_1})$ from $\bboard$, accepts the sender's payment if $\validity(VK_{\cbank},\msgs_{valid},\token^{\sender_1,\receiver_1})=1$. Otherwise, if receiving $(\denied,\vkso,\vkro)$ from $\sender_1$, rejects the payment.
 \end{itemize}
      \item[Token Validation:] Each user $\user\in\users$,
     when a new asset $F$ appears in $\msgs$, checks that $\validity(VK_{\cbank},\msgs_{valid}^\user,F)=1$. If the check passes, sets $\verassets^\user\gets\verassets^\user\cup\{F\}$.
\end{description}

\begin{figure}[htbp]
\begin{framed}
    Takes as inputs a set of users $\users$, banks $\banks\subseteq\users$, and the central bank $\cbank$. Let $\msgspace=\bin^*$ be the space of the messages accepted by the bulletin board. $\bboard$ initializes a set of messages $M\gets\emptyset$ and an index $j=0$. It behaves as follows:
    \begin{itemize}
        \item When receiving the command $(\post,\msg)$ where $\msg\in\msgspace$ from a bank $\bank\in\banks$ or from $\cbank$,  set $\msgs\gets \msgs\cup\{(j,\msg)\}$, set $j=j+1$ and send $(\stored,\msg)$ to each $\user\in\users$.
        \item When receiving the command $(\readcom)$ from a user $\user\in\users$, send $\msgs$ to $\users$.
    \end{itemize}

\end{framed}
    \caption{Bulletin Board $BB$}
    \label{fig:funcbb}
\end{figure}

\begin{figure}[htbp]
\begin{framed}
   Takes as inputs a set of users $\users$ and a subset of users $\banks\subseteq\users$ and a value $\tau$ indicating the number of assets each bank should receive and hold. The setup works as follows:
    \begin{description}

                \item[Token generation:] The following steps are performed:
                \begin{itemize}
                     \item Compute $(\vkcb^{(\bank,i)},\skcb^{(\bank,i)})\getsr\SIG.\Gen(1^\secpar)$ for each $i\in [\tau]$ and each $\bank\in\banks$.
                    \item Wait to receive $\{\vk_i^\bank\}_{i\in[\tau]}$ from each $\bank\in\banks$.  
                    \item Finally, compute $\sigma^{(\bank,i)}_{\cbank}\getsr\SIG.\sign(\skcb^{(\bank,i)},\vk_i^\bank)$ for each $i\in [\tau]$.
                \end{itemize}
                \item[CRS generation:]
                Compute $(\vk_\crs,\sk_\crs)\getsr\SIG.\Gen(1^\secpar)$, $\crs\getsr\nizk.\setup(1^\secpar)$ and  $\sigma^\crs_{\cbank}\getsr\SIG.\sign(\sk_\crs,\crs)$.
        \item[Token distribution:] 
    Distribute central bank verification keys $\{\vkcb^{(\bank,i)}\}_{i\in[\tau],\bank\in\banks}$ and the CRS key $\vk_\crs$ to each $\user\in\users$ in an authenticated fashion.
   Send $(\post,(\crs,\sigma^\crs_{\cbank}))$ to $\bboard$, and send $(\post,(\vkcb^{(\bank,i)},\vk_i^\bank,\sigma^{(\bank,i)}_{\cbank}))$ to $\bboard$ for each $i\in [\tau]$ and each $\bank\in\banks$. 
               \end{description}

\end{framed}
    \caption{Central Bank Setup Protocol}
    \label{fig:funccb}
\end{figure}

\section{Proofs}\label{s:proofs}
\subsection{Proof of Claim~\ref{claim:tokenunf}}\label{app:tokenunf-proof}

Recall that the common random string is generated honestly by the challenger (acting as a central bank) during setup.
Also, recall that the honest players signing keys as well as the secrets of the genesis tokens $VK_{\cbank}$ are generated by the challenger/oracle but kept oblivious to the adversary.

The adversary, to forge a new token maliciously,
shall be able to generate a new valid token $\token$ without redeeming a live token they own, i.e. for which they own the signing key corresponding to the receiver verification key of such a live token. In particular, given the most updated value $\msgs_{valid}$, the output $\validity(VK_{\cbank},\msgs_{valid},F)$ shall be $1$.
We distinguish between the following cases:
\begin{enumerate}
    \item\label{it:proofgen} Generate new proof for a token the adversary does not own, i.e. for which he does not know the signing key related to the receiver verification key.
    \item \label{it:proofreuse} Re-use the same proof of another token (could be an adversarial or an honest one).
    \item\label{it:proofspend} Generate a new proof related to a token that has been already burnt and then redeemed by the adversary (double-spending).
\end{enumerate}
Regarding case \ref{it:proofgen}, we distinguish three sub-cases.

First, the adversary might try to burn a live token for which he does not know the signing key corresponding to the receiver verification key and then use the commitment randomness used to produce the burning factor of such burnt token as a witness for the ZK-PoK $\pi$ that will be attached to the new token $\token$.
However,
 we recall that each burnt token must be accompanied by a signature $\sigma^\receiver$ of a message containing the sender verification key $\vk^\sender$ together the burning factor $\blinded^\receiver$ and the signature should verify verifying under the receiver verification key $\vk^\receiver$, i.e. $\SIG.\ver(\vk^\receiver,(\vk^\sender,\blinded^\receiver),\sigma^\receiver)=1$.
 Such a token will not be ever added by any honest receiver to its set of valid token $\msgs_{valid}$ if the signature is incorrect ({\em Setup} and {\em Token Validation} phases) since this check is performed when running $\validity(VK_{\cbank},\msgs_{valid},F)$.
 Thanks to the EUF-CMA property of $\SIG$, an adversary cannot forge a valid signature $\sigma^\receiver$ such that $\SIG.\ver(\vk^\receiver,(\vk^\sender,\blinded^\receiver),\sigma^\receiver)=1$ with non-negligible probability.

Alternatively, the adversary, given any adversarially chosen subset of valid tokens in $\msgs_{valid}$, might try to find an opening $(\vk,r)$ for one of such burning factors, allowing him to produce an accepting ZK-PoK with $\vk$ as part of the statement $r$ as part of the witness. However, this would break the binding property of the underlying commitment scheme with non-negligible probability.

Finally, the adversary might try to forge an accepting proof $\pi$ for the relation $\Reldc^\Commit$ and statement $(\blinded_1,\ldots,\blinded_n,\vk)$ where $\blinded_1,\ldots,\blinded_n$ is the set of burning factors. However, for the Knowledge Soundness of $\nizk$, we would be able to invoke the underlying extractor 
to extract the index $i$ and the opening randomness $r$ from the adversary

Regarding case \ref{it:proofreuse}, notice that the verification key is part of the statement. The $\validity$ algorithm forbids the re-use of a sender verification key. Hence, no accepting ZK proof for a re-used sender verification key will ever be accepted.

Regarding case \ref{it:proofspend}, similarly to case \ref{it:proofreuse}, if the adversary had already spent their burnt token, the verification key embedded inside the burning factor of the adversary's burnt token had already appeared in $\msgs_{valid}$. Hence, since the adversary must produce a proof having as a statement, an already appeared verification key and since the  $\validity$ algorithm forbids the re-use of a sender verification key, such a new token will not be ever accepted or added into $\msgs_{valid}$ by the honest players.

To conclude the proof, we argue that the genesis tokens generation cannot be performed by the adversary since the sender verification keys of the central bank are generated and certified by the central key itself (impersonated by the challenger at the beginning). Due to the EUF-CMA property of $\SIG$, the adversary cannot generate genesis tokens and certify them with a signature verifiable under the central bank verification key $\vk_{\cbank}$.
\subsection{Proof of Claim~\ref{claim:tokenind}}\label{app:tokenind-proof}

    Notice that an adversary is not able guess the correct bit if the game in which the challenger's choice is fixed to $b$ is indistinguishable, in the eyes of the adversary, from the game in which the challenger's choice is fixed to $1-b$.

        Let $\msgs_{burnt}$ be the set of burnt tokens with honest receiver.
    Let $\msgs^*_{burnt}$ be the set of burnt tokens with honest sender and receiver.

    Let us consider a sequence of hybrid arguments as follows. Let $\gamma=|\msgs^*_{burnt}|$ and $\tau=|\msgs_{burnt}|$.
    \begin{itemize}
        \item $\hyb^b_0$: Identical to the original game, except that the challenger's bit choice is fixed to $b$.
         \item $\hyb^{1-b}_0$: Identical to the original game, except that the challenger's bit choice is fixed to $1-b$.
        \item $\hyb^b_i, 0\le i\le\gamma$:
       Identical to the original game, except that, in the first $i$ live tokens whose sender and receiver are honest, NIZK proofs are replaced with simulated proofs, i.e. in each $F_k=(\vk^\sender_k,\vk^\receiver_k,\sigma^\sender_k,\indexes_k,\pi_k)\in \msgs^*$ with $k\le i$ of $\hyb_0^b$, $\pi_k$ is replaced with $\Sim(\td,(\blinded_1,\ldots,\blinded_n,\vk^\sender))$, where $\blinded_j$ is the burning factor of $\restrict{\msgs_{burnt}}{\indexes}[j]$.
       \item $\hyb^b_{\gamma+\ell},0 \le\ell\le\tau$: Identical to $\hyb^b_{\gamma}$, except that the first $\ell$ burning factors in $\msgs_{burnt}[\ell]$ are replaced with $\commit(0;r)$ for a random $r\gets\bin^\secpar$.
        \item $\hyb^{1-b}_{\gamma+\ell},0 \le\ell\le\tau$: Identical to $\hyb^{1-b}_{\gamma}$, except that the first $\ell$ burning factors in $\msgs_{burnt}[\ell]$ are replaced with $\commit(0;r)$ for a random $r\gets\bin^\secpar$.
       \item $\hyb^{1-b}_i, 0\le i\le\gamma$: Identical to the original game, except that, in the first $i$ live tokens whose sender is an honest user,  NIZK proofs are replaced with simulated proofs, i.e. in each $F_k=(\vk^\sender_k,\vk^\receiver_k,\sigma^\sender_k,\indexes_k,\pi_k)\in \msgs_{send}$ with $k\le i$ of $\hyb_0^{1-b}$, $\pi_k$ is replaced by $\Sim(\td,(\blinded_1,\ldots,\blinded_n,\vk^\sender))$, where $\blinded_j$ is the burning factor of $\restrict{\msgs_{burnt}}{\indexes}[j]$.

    \end{itemize}

Our aim is to show that $\hyb_0^b\approx_s\hyb_0^{1-b}$.
Let us show the following:
\begin{itemize}
    \item $\hyb^b_i \approx_s \hyb^b_{i+1}$ for $0\le i < \gamma$:
    The two hybrids are indistinguishable thanks to a straightforward reduction to statistical zero-knowledge of $\nizk$. Indeed, the only difference  is that in $\hyb^b_{i}$  the $i$-th proof is simulated whereas in $\hyb^b_{i+1}$ the proof is honestly generated. Since the sender is honest, the adversary does not have knowledge of the random coins used to generated such proofs in order to distinguish.
    \item $\hyb^b_{\gamma+\ell}\approx_s \hyb^b_{\gamma+\ell+1}$ for $0\le\ell<\tau$:
    The two hybrids are statistically close thanks to the statically hiding property of $\com$. Indeed, the only difference  is that in $\hyb^b_{\gamma+\ell}$  the $\ell$-th burning factor of $\msgs_{burnt}$ is a commitment of a verification key, whereas in $\hyb^b_{i+1}$ it is a commitment that is independent of such verification key (commitment of $0$). Since the receiver is honest, the adversary does not have knowledge of the random coins used to generated such commitments in order to distinguish.
    \item $\hyb^{1-b}_i \approx_s \hyb^{1-b}_{i+1}$ for $0\le i < \gamma$ and $\hyb^{1-b}_{\gamma+\ell}\approx_s \hyb^{1-b}_{\gamma+\ell+1}$ for $0\le\ell<\tau$: Same reasons used for $\hyb^{b}_i$ and $\hyb^{b}_{\gamma+\ell}$ apply.
    \item $\hyb^b_{\gamma+\tau}\equiv\hyb^{1-b}_{\gamma+\tau}$: The difference lies in the two hybrids lies in its payment trace between honest users and in the burning factors of token in which the sender is adversarial.
    Notice that proofs and burning factors of the honest trace are replaced with simulated proofs and commitment of $0$, respectively.
    Also, the burning factors of honest receivers in the adversarial trace are replaced with the commitment of $0$.
    Hence, due two hybrids have the same distribution.

\end{itemize}
The proof follows since $\hyb^b \approx_s \ldots \approx_s \hyb^b_i \approx_s\ldots\approx_s \hyb^b_{\gamma} \approx_s \ldots \approx_s \hyb^b_{\gamma+\ell} \approx_s \ldots\hyb^b_{\gamma+\tau}\equiv\hyb^{1-b}_{\gamma+\tau} \approx_s \ldots\approx_s \hyb^{1-b}_{\gamma+\ell} \approx_s \ldots\approx_s \hyb^{1-b}_{\gamma}\approx_s \ldots\approx_s \hyb^b_i\approx_s\ldots\approx_s\hyb^{1-b}$.
\section{A Faster Accountability-Based Alternative}
\label{s:alternative}

As already pointed out, if we do not want to rely on parties with enhanced computing capabilities to calculate our recursive zk-SNARKS, we require that all the players must take into their memory the entire set of tokens to verify the ZK proofs attached to the token transfer. Other than the memory requirement, the computational burden is problematic too. Indeed, each time a new proof appears, it must be verified by each wannabe receiver. 

In the following, we propose an alternative version of our protocol where a new semi-trusted party, which we call {\em minter}, takes the burden of validating new tokens, hence eliminating the need of providing a NIZK-PoK each time a new transaction is done.
When a minter misbehavior is observed, the central bank could kick out the minter from the system and execute a fail-safe mechanism.
Here, we admit the usage NIZK-PoKs from the sender who transferred tokens through the corrupted minter to reduce the number of tokens that are going to be discarded during the fail-safe mechanism. 

\paragraph{Our Minter-aided protocol.} Differently from our main protocol $\pidc$, the tokens generated as an outcome of a new transaction are not validated directly by the sender through a NIZK-PoK but by a new entity called minter, whose aim is to ``mint" these coins via a blind signature of the sender verification key.
By introducing the minter in the system, we need to relax the security requirements. We now require full token privacy (as in \cref{s:analysis}) together with a relaxed token integrity property, which we call minter accountability, stating that when a minter creates a new token from scratch or double-spends an existing token, they will be eventually spotted.
The protocol works as follows:
\begin{description}
    \item[Setup:] During setup, each minter $\minter$ generates a pair $(\vk^\minter,\sk^\minter)$ send sends $\vk^\minter$ to $\cbank$, to be posted by the latter into $\bboard$. 
    \item[Burning:] As in $\pidc$ the receiver $\receiver_0=\sender_1$ of a token $F^{\sender_0,\receiver_0}=(\vksz,\vkrz,\sigsz,\allowbreak\sigma^{\minter_0})$ burns $F^{\sender_0,\receiver_0}$ by adding to it a burning factor $\blinded$. Such a factor is computed by the sender with the $\blind$ algorithm of $\BS$, i.e. $\blinded=\blind(\vkso;r)$, where $(\vkso,\skso)\getsr\BS.\kgen(\secpar)$.
    \item[Spending:] Differently from to $\pidc$, the sender $\sender_1$, upon receiving $\vkro$ from $\receiver_1$, blindly ask for a signature of the message $(\vkro)$ to a minter $\minter$, obtaining $\sigma^\minter$. More precisely, the interaction between $\sender_1$ and $\minter$ works as follows:
    \begin{itemize}
        \item $\sender_1$ sends $\blinded$ to $\minter$.
        \item $\minter$ computes $\tilde{\sigma}\getsr\BS.\bsign(\sk^\minter,\blinded)$ and sends $\tilde{\sigma}$ to $\sender_1$.
        \item $\sender_1$ computes $\sigma^\minter=\unblind(\blinded,\tilde{\sigma},r)$.
    \end{itemize}
   Finally, $\sender_1$ signs $\vkro$ with $\skso$ obtaining $\sigso$, i.e. $\sigso\getsr\SIG.\sign(\skso,\vkro)$ and sends a new live token  $F^{\sender_1,\receiver_1}=(\vkso,\vkro,\sigso,\sigma^\minter)$ to $\bboard$.

\end{description}
 Since $\BS.\ver(\vk^\minter,\vkro,\sigma^\minter)=1$ and $\SIG.\ver(\vkso,\vkro,\sigso)=1$, the token is authorized by the sender (because of $\sigso$) and minted by $\minter$ (because of $\sigma^\minter$).

  Tokens generated from the same chain of payment (initially started from a Genesis token released by the central bank as in $\pidc$) are always minted the same minter. It means that when an old token is burned in favor of a new token, the minter signing the new token should be the same one who signed the burnt one.
We may alternatively consider a scenario in which a different minter can mint a token of a different chain of payment via a pre-determined agreement between the two minters. A token generated by a different minter without an agreement beforehand is considered invalid.
 
The blindness of $\BS$ guarantees token privacy of the protocol. We will now briefly discuss our fail-safe mechanism to ensure minter accountability.

\paragraph{Accountability mechanism against minters corruption.}
All the minters have a fixed amount of tokens that are allowed to mint (i.e., sign blindly).

When the minter burns all of their allocated mints (i.e., minter signatures), the central bank validates a new minting key provided by the minter and allows the latter to restart minting.

If the minter mints more than expected, it will be flagged as corrupted by the users/central bank, and all the tokens minted by the corrupted minter will be flagged as forged.

If an honest user wants to keep their minted token, they can produce a NIZK-PoK for the relation $\Reldc^\blind$ showing that their verification key was blinded (through $\blind$ instead of $\commit$) inside a burning factor of a live valid token. Here, the token validity predicate further checks that the token is not flagged as forged.
Alternatively, an honest user not caring about token privacy can directly open the burning factor $\beta$ connected to a burnt token by releasing the random factor provided during the blinding procedure.

Notice that such a NIZK-PoK could be generated, when interacting with an
honest minter, by users who choose to do so.  This NIZK-PoK can then be
furnished to recipients, who can in turn provide added value to minters by
sharing it as evidence that can be used to programmatically allow the minter to
mint extra tokens without immediately asking the central bank for a new minting
key.

\end{document}